\documentclass[prb,preprint]{revtex4-1} 

\usepackage{amsmath}

\usepackage{graphicx}

\usepackage{mathtools}

\begin{document}

\title{Understanding Current Signals Induced by Drifting Electrons}

\author{Kristen A. Recine}
\affiliation{Bryn Mawr College, Department of Physics, Bryn Mawr, PA 19010, USA}
\email{krecine@brynmawr.edu}
\author{James B. R. Battat}
\affiliation{Wellesley College, Department of Physics, Wellesley, MA, 02481, USA}
\author{Shawn Henderson}
\affiliation{Department of Physics, Cornell University, Ithaca, NY 14853, USA}
\date{\today}

\begin{abstract}

Consider an electron drifting in a gas toward a collection electrode. ÊA common misconception is that the electron produces a detectable signal only upon arrival at the electrode. ÊIn fact, the situation is quite the opposite. ÊThe electron induces a detectable current in the electrode as soon as it starts moving through the gas. ÊThis induced current vanishes when the electron arrives at the plate. To illustrate this phenomenon experimentally, we use a gas-filled parallel plate ionization chamber and a collimated $^{241}$Am alpha source, which produces a track of a fixed number of ionization electrons at a constant distance from the collection electrode. We find that the detected signal from the ionization chamber grows with the electron drift distance, as predicted by the model of charge induction, and in conflict with the idea that electrons are detectable upon arrival at the collection plate.

\end{abstract}

\maketitle

\section{Introduction}
\label{sec:intro}
Consider an electron drifting towards an electrode. At what point does an electronic signal appear on that electrode? Is it when the electron arrives at that electrode, or when the electron begins to move? Despite the frequent use of the term ``charge collection,'' an induced current in the electrode can be measured as soon as the electron begins to move. Although the phenomenon of charge induction is well--described in the literature,~\cite{spieler}$^{,}$~\cite{knoll}$^{,}$~\cite{rossi} a surprisingly common misconception is that no detectable signal exists until the electron arrives at the electrode.

\begin{figure}[tb]
\centering
\includegraphics[width=6.0 in]{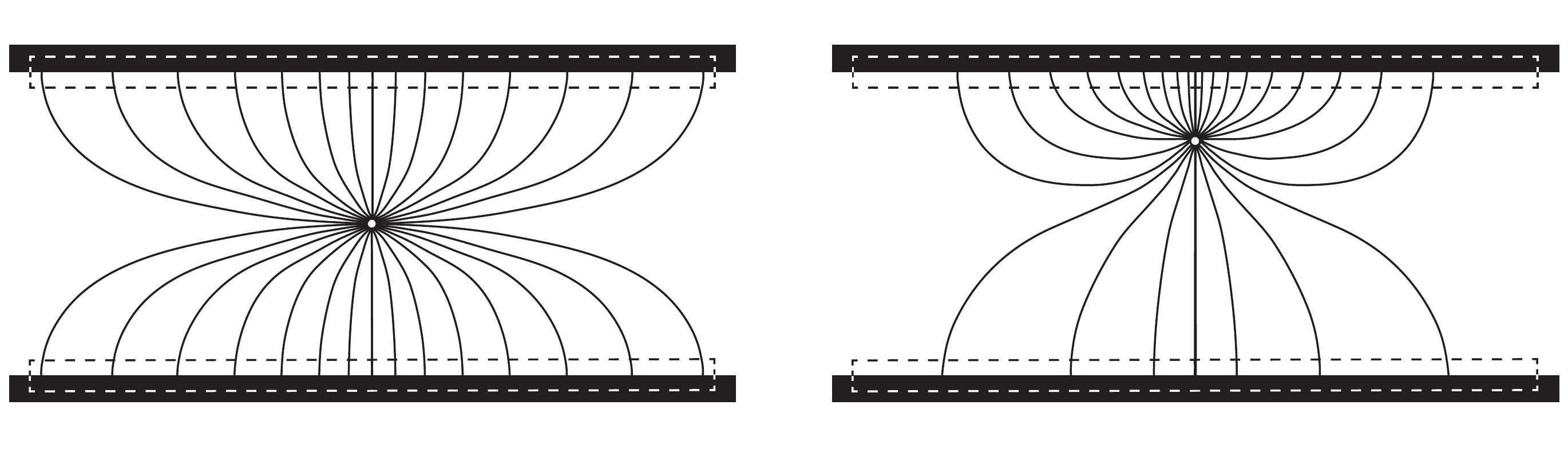}
\caption{Left: An electron sits halfway between two grounded electrodes and an equal number of electric field lines terminate on each electrode. Right: The electron is moved closer to the top plate and therefore more electric field lines terminate on the top plate than on the bottom plate. The dotted lines indicate the Gaussian surface used in Gauss's law.}
\label{fig:fieldlines}
\end{figure}

Following the discussion presented in Ref.~\onlinecite{spieler}, the process by which an electronic signal is produced when the electron moves can be understood by considering an electron that is situated midway between two parallel grounded conducting plates, as shown in Fig.~\ref{fig:fieldlines}. Symmetry requires that half of the electric field lines terminate on each plate, so half of the total electron charge will be induced on each plate. If the electron then shifts towards the top plate, more field lines will terminate on that plate. Now consider Gauss's law for a rectangular Gaussian surface at the top plate (see Fig.~\ref{fig:fieldlines}),
\begin{equation}
\label{eq:gauss}
\oint_{S}\vec{E}\cdot d\vec{a}=\frac{Q_{enc}}{\epsilon_{0}},
\end{equation}
in which $\vec{E}$ is the electric field, $d\vec{a}$ is the normal area vector, and $Q_{enc}$ is the total charge enclosed in the Gaussian surface. An increase in the electric flux corresponds to an increase in the enclosed charge. This change in induced charge over time represents a current. Thus the motion of the electron induces a current in the top plate, even though the electron has not yet reached the plate.~\cite{footnote0} The induced current in the electrode results from the changing number of electric field lines that terminate on the electrode and does not depend on the amount of charge that reaches the electrode.

In this paper we present an experiment conducted in an ionization chamber that demonstrates this aspect of charge induction. By placing a collimated alpha source on a remotely adjustable platform we are able to generate ionization electrons in the chamber and control the total distance over which those ionization electrons drift. We are able to therefore study the effects of drift distance (and readout electronics) on the measured induced charge signal without otherwise disturbing the setup by opening the ionization chamber. We analyze the average voltage pulse amplitude for different source positions (electron drift distances) and show that, for the same average number of electrons produced at each source position, electrons traveling longer distances produce larger measurable signals. Thus, the total induced charge depends on the drift distance of the electrons. 

This setup can be used to localize the position of ionization events inside the detector and estimate the electron drift speed in the detector. We also show that the readout electronics (used to integrate and shape the output signal) can diminish the amplitude of the observed output voltage pulse if the total drift time of the electrons is comparable to the decay timescale of the charge integrating electronics. This experiment can be performed with equipment that is commonly used in advanced undergraduate laboratories, and is appropriate for such students. Although ionization chambers were introduced in the early 1900s,~\cite{flakus} they are still used in particle and nuclear physics today. In particular, the type of measurements described in this paper is actively being explored as methods for rejecting radioactive backgrounds in directional dark matter experiments.~\cite{whitepaper}$^{,}$~\cite{drift}

This paper is outlined as follows: Section~\ref{sec:ioncharge} reviews the ionization chamber and the physics behind charge induction. In Section~\ref{sec:setup}, we describe the experimental setup, including the remotely adjustable collimated alpha source. The measurements obtained with this setup are presented and analyzed in Section~\ref{sec:results}. Finally, we draw conclusions and describe possible extensions of this work in Section~\ref{sec:conclusions}. 

\section{Signal Production and Measurement}
\label{sec:ioncharge}
\subsection{Ionization}
\label{sec:ion}
When a charged particle such as an electron, ion, or alpha particle moves through a medium, it loses energy through interactions with molecules in that medium.~\cite{ahlen}$^{,}$~\cite{pdg} In this work, we arrange for an alpha particle to lose energy through collisions with a gas in an ionization chamber. In this scenario, we will consider the case in which the kinetic energy of the alpha particle is lost to ionization of the surrounding gas through Coulomb interactions.~\cite{footnote1} The resulting electron--ion pairs will rapidly recombine unless an external drift electric field (hereafter referred to as the drift field) is applied, in which case the electrons and ions drift apart. The signal resulting from the moving charges provides information on the energy and position of the original alpha particle.~\cite{knoll}
\begin{figure}[tb]
\centering
\includegraphics[width=3.4 in]{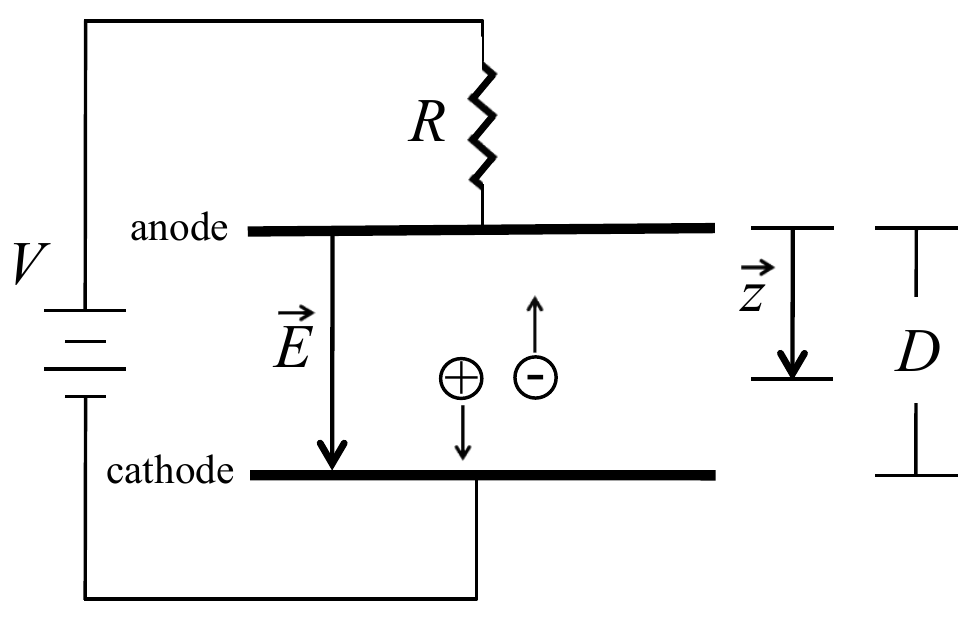}
\caption{Schematic of an ionization chamber. The rectangular electrodes are viewed edge--on. A single electron--ion pair has been produced at a distance $z$ from the anode.}
\label{fig:ionizationchamber}
\end{figure}

Figure~\ref{fig:ionizationchamber} shows a schematic of a parallel plate ionization chamber, which consists of two flat, rectangular parallel electrodes separated by some distance $D$. We will assume that $D$ is small compared to both the length and width of the electrodes so that the drift field $\vec{E}$ is uniform and normal to the electrodes, with magnitude
\begin{equation}
\label{eq:efield}
E=\frac{V}{D}.
\end{equation}
The plates are placed in a gas at pressure $P$ and support a potential difference $V$. An ionization electron will travel a distance $z$ to the anode, so a positive ion travels a distance $(D-z)$ to the cathode. 

Ionizing radiation with kinetic energy $K$ produces, on average, $N_{e}=\dfrac{K}{W}$ electron--ion pairs, where $W$ is the $W$--value of the gas (which defines the average energy required to liberate an electron--ion pair in the gas). Note that the $W$-value is always larger than the ionization energy, because it is not true (in general) that all of the ionizing particle's kinetic energy generates ionization.~\cite{knoll} Typical $W$--values for alpha particles in commonly used detector gases are 20--40~eV/electron-ion pair.~\cite{blum} Our chamber is filled with P--10 (90\%~Ar~+~10\%~CH$_{4}$ by volume), a common detector gas, though many other gases could be used. P--10 has a $W$--value of~26~eV/electron--ion pair.~\cite{knoll} Since alpha particles produced from radioactive decays typically have energies of 4--7 MeV, an alpha particle that fully stops in the gas will produce a few hundred thousand primary electron--ion pairs that can be detected without the need for gas amplification.

After primary ionization occurs in a gas, it is possible for the freed electrons to gain enough energy to produce secondary ionization. In order for this to happen, the electron energy must exceed the ionization potential of the gas. This is possible if the drift field can sufficiently accelerate the primary electrons. By carefully operating at low drift fields and high pressure, we ensure that we are well outside the secondary ionization regime for P--10 gas, and that we are working in the limit of no gas amplification (the first Townsend coefficient is zero), as verified by a MAGBOLTZ~\cite{magboltz} simulation.

\subsection{Drift Velocity}
\label{sec:drift}
As stated above, the electrons and ions created during ionization will drift towards opposite plates in the ionization chamber due to the drift field. The drift velocity depends on the drift field $\vec{E}$, and the mobility $\mu$. The mobility depends on the particle and the gas through which the particle moves, and is inversely proportional to the pressure, $P$. The drift velocity for electrons is given by~\cite{rossi}
\begin{equation}
\label{eq:ve}
\vec{v}_{e}=-\mu_{e}\vec{E},
\end{equation}
and similarly for ions:
\begin{equation}
\label{eq:vi}
\vec{v}_{i}=\mu_{i}\vec{E}.
\end{equation}

Electrons move through P--10 gas over 1000 times faster than ions because the electrons are less massive and have a longer mean free path.~\cite{yamashita} The drift field accelerates the electrons, but collisions with the surrounding gas molecules retard the motion, with the net result being that the electrons drift at a constant speed that depends on $E/P$. For a given value of $E/P$, the drift time $t_{e}$ for an electron to travel to the anode depends linearly on $z$ as
\begin{equation}
\label{eq:te}
t_{e}=\frac{z}{v_{e}},
\end{equation}
while the ions reach the cathode in a time $t_{i}$:
\begin{equation}
\label{eq:ti}
t_{i}=\frac{D-z}{v_{i}}.
\end{equation}

\subsection{The Induced Charge}
\label{sec:ideal}
The current $i$ that a moving charge $q$ induces on a specific electrode can be calculated from the Shockley--Ramo Theorem,~\cite{blum}$^{,}$~\cite{shockley}$^{,}$~\cite{ramo} giving
\begin{equation}
\label{eq:ramo}
i=\dfrac{-q}{V_{Q}}\,\vec{v}_{q} \cdot \vec{E}_{Q},
\end{equation}
where $\vec{v}_{q}$ is the velocity of the moving charge and $V_{Q}$ is the weighting potential applied to the electrode in question to create the so--called weighting field $\vec{E}_{Q}$. The weighting field is the hypothetical drift field obtained when the electrode of interest is held at unit potential ($V_{Q}$~=~1~V) and all others are grounded. In our case, the cathode is grounded and we measure the induced current on the anode. In general, the weighting field is different from the drift field; however, for a two--electrode parallel plate ionization chamber, they are equivalent [given in Eq.~\eqref{eq:efield}]. In order to calculate the current induced on the anode by the drifting electrons, we first note that the weighting field is $\vec{E}_{Q}=\dfrac{1\,\text{Volt}}{D}\,\hat{z}$. Recalling that the field points opposite to the electron drift velocity, we have $\vec{v}_{e}\cdot\vec{E}_{Q}=-\dfrac{v_{e}}{D}$, and the induced current on the anode due to a single drifting electron is $-\dfrac{e\,v_{e}}{D}$. This result is then scaled by the number of drifting electrons, so for $N_{e}$ electrons produced at the same distance $z$ from the anode, the induced current is~\cite{ramo}
\begin{equation}
\label{eq:ie1}
i_{e}(t)=\begin{cases}
	-N_{e}\dfrac{e\,v_{e}}{D} & 0 < t \le t_{e} \\
	0 & t > t_{e}
\end{cases}
\end{equation}

Equation~\eqref{eq:ie1} states that the induced current in an ionization chamber is constant in time for as long as the electrons are drifting (again, assuming a common drift distance and time for all electrons). Once an electron reaches the anode it no longer induces a current on the anode, so $i_{e}=0$. The integral of $i_{e}(t)$ over time, which we denote as $Q_{e}$, represents the total change in induced charge on the anode due to the $N_{e}$ drifting electrons. Assuming that all electrons are produced at the same distance $z$ from the anode,
\begin{equation}
\label{eq:qinde}
Q_{e}=\int i_{e}dt = i_{e}t_{e}=-N_{e}e\frac{z}{D}.
\end{equation}
Similarly, the induced charge $Q_{i}$ on the anode from the drifting ions is
\begin{equation}
\label{eq:qindi}
Q_{i}=\int i_{i}dt = i_{i}t_{i}=-N_{e}e\left(1-\frac{z}{D}\right),
\end{equation}
where we have assumed that the ions are singly ionized so that $q_{i} = -q_{e}$ and $N_{e}=N_{i}$.

We see from Eq.~\eqref{eq:qinde} that the induced charge from the electrons depends on both the number of drifting electrons and on the fraction of the plate separation over which the electrons drift. The maximum possible charge induced by the electrons is $-N_{e}e$ and is achieved when the electrons drift the full length of the gap $D$. If the electrons are produced halfway between the plates, then $Q_{e}=-\dfrac{N_{e}e}{2}$, and so on. Combining this expression for $Q_{e}$ with Eq.~\ref{eq:qindi}, the total induced charge on the anode is then $Q_{total}=Q_{e}+Q_{i}=-N_{e}e$, which is independent of the drift distance. However, as we now describe, the highly disparate drift speeds of electrons and ions ensure that the detector readout electronics will not be equally sensitive to the electron and ion signals. As a result, the electronics introduce a $z$--dependent system response, as desired for this work.

\subsection{The Effect of Charge Readout Electronics on the Measured Induced Charge}
\label{sec:nonideal}
\begin{figure}[tb]
\centering
\includegraphics[width=3.4 in]{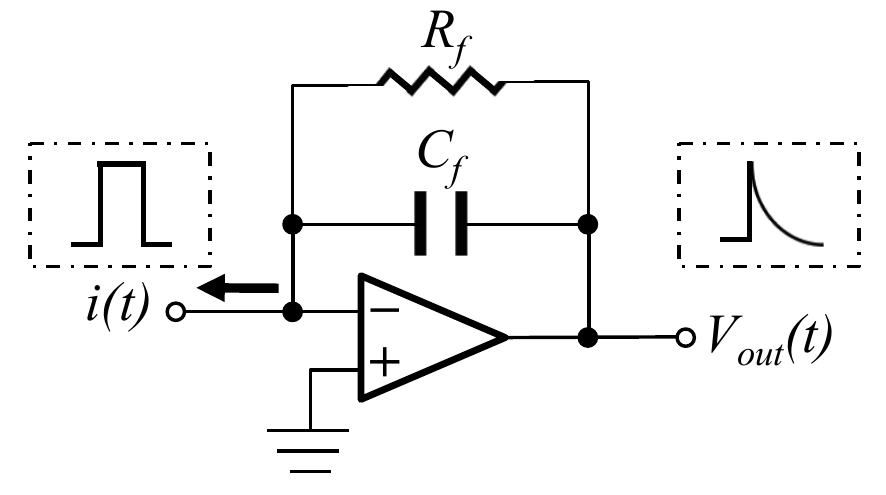}
\caption{Basic charge--integrating amplifier with feedback capacitance $C_{f}$ and resistance $R_{f}$. A fast current pulse $i(t)$ at the input is integrated on the capacitor to produce a voltage pulse at the output with decay timescale $\tau_{p}=R_{f}C_{f}$.} 
\label{fig:basicpreamp}
\end{figure}

The induced charge signal derived above does not take into account any electronic apparatus that may be connected to the detector. Charge--integrating amplifiers, commonly known as preamplifiers, are often used to produce an output voltage pulse whose amplitude is proportional to the integral of the input current pulse.~\cite{knoll} Figure~\ref{fig:basicpreamp} shows a basic preamplifier circuit diagram, which consists of an operational amplifier, a feedback resistor with resistance $R_{f}$, and a feedback capacitor of capacitance $C_{f}$. The capacitor integrates the input current pulse while the resistor provides a discharge path for the capacitor with a discharge timescale $\tau_{p}=R_{f}C_{f}$. In the absence of the resistor, the input current flows onto the capacitor such that the output voltage is $V_{out}=-\left(\dfrac{1}{C_{f}}\right)\displaystyle\int{i(t)dt}$. The resistor discharges the stored charge in preparation for the next event in the detector.

If the preamplifier were equally sensitive to the charge induced by the electrons and the ions, then the amplitude of the output voltage pulse would be proportional to $Q_{total}$ and therefore independent of the electron drift distance $z$ (as stated in Section~\ref{sec:ideal}). However, because the ion drift time is long compared to the discharge timescale of the preamplifier, it is not true that the total induced charge $Q_{total}$ is equal to the sum of $Q_{e}$ and $Q_{i}$. If the duration of the current pulse is on the order of the decay time $\tau_{p}$, some of the charge stored on the capacitor will discharge through the resistor before the current pulse terminates. This reduces the amplitude of the output pulse. Commercially available preamplifiers typically have $\tau_{p}\approx$~10--100~$\mu s$. The electron drift speed depends on the drift field but a representative value for this experiment is 0.25~cm/$\mu$s,~\cite{yamashita} and so the total electron drift time across the full electrode separation (5.08~cm) is 19~$\mu$s. The ion drift speed, on the other hand, is a factor of 1000~smaller, and so $t_{i}$/$\tau_{p}\gg100$ even for the shortest ion drift time expected in this experiment ($E/P$~=~22~V~cm$^{-1}$~atm$^{-1}$ and $1-(z/D)$~=~0.2). The induced current from drifting ions therefore makes a negligible contribution to the total induced current, and the preamplifier output amplitude is approximately proportional to $Q_{e}$ only and will therefore depend on the drift distance, as seen in Eq.~\eqref{eq:qinde}.

A detector designed to measure the total deposited energy due to ionization must avoid any $z$--dependence in the measured energy created by the decay time, and this is typically done through the use of a Frisch grid.~\cite{knoll} In our work, however, this position dependence is desired so no Frisch grid was used.

In order to demonstrate how the electron drift distance $z$ affects the amplitude of the output voltage pulse, we now calculate the output voltage $V_{out}(t)$ for a given input current $i(t)$. This can be done by convolving the input current with the impulse response of the preamplifier. The impulse response is the temporal response of the preamplifier to a delta function current pulse at the input. The impulse response $h(t)$ for a charge--integrating preamplifier is~\cite{blum}
\begin{equation}
\label{eq:transfer}
h(t)=Ae^{-t/\tau_{p}} \quad \text{for} \,\,\, t \ge 0,
\end{equation} 
where $A=\dfrac{1}{C_{f}}$ is the preamplifier conversion gain from charge to volts (e.g., in mV per million electron--ion pairs or V per pC). The output voltage from the preamplifier is then the convolution of $i(t)$ and $h(t)$, or
\begin{equation}
\label{eq:voutt}
V_{out}(t)=i(t) \otimes h(t)=\int^{\infty}_{-\infty} i(t-t')h(t') dt',
\end{equation}
where the current is given by Eq.~\eqref{eq:ie1} because $t_{e}<\tau_{p}\ll t_{i}$ and so we take $i(t)\approx i_{e}(t)$. We are interested in the $z$--dependence of the output voltage amplitude peak which occurs at $t=t_{e}$. The expression for the peak output voltage becomes
\begin{equation}
\label{voutte}
V_{peak}=V_{out}(t_{e})=A\tau_{p}\frac{N_{e}e\,v_{e}}{D}\left(1-e^{-t_{e}/\tau_{p}} \right).
\end{equation}
Using Eq.~\eqref{eq:te}, we can write this expression as a function of the fractional drift distance $z/D$ (left intentionally in the exponential) as
\begin{equation}
\label{eq:voutxD}
V_{peak}\left(\frac{z}{D}\right) =AN_{e}e\frac{v_{e}\tau_{p}}{D}\left(1-e^{\frac{-z}{D}\frac{D}{v_{e}\tau_{p}}}\right).
\end{equation}

At any given time, $V_{out}$ is equal to the charge stored on the feedback capacitor divided by the capacitance $C_{f}$, so we can compare Eq.~\eqref{eq:qinde} with Eq.~\eqref{eq:voutxD}. In both cases, we see that the output voltage signal depends on the drift distance of the electrons. In addition, notice that to first order in $t_{e}/\tau_{p}$, Eq.~\eqref{eq:voutxD} reduces to Eq.~\eqref{eq:qinde}. In other words, when the electron drift time is small compared to the preamplifier decay time, the full induced electron charge signal is preserved and the readout electronics do not distort the electron--induced current. For the same reason, the long ion drift time ensures that the ion--induced current contributes negligibly to the output voltage.

\begin{figure}[tb]
\centering
\includegraphics[width=3.4 in]{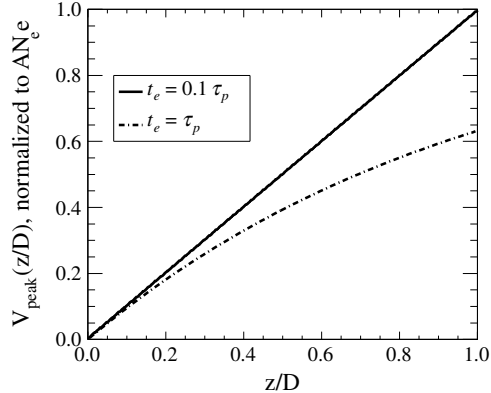}
\caption{Plot of Eq.~\eqref{eq:voutxD} for two separate current pulse durations to show their effect on the output voltage. Here, $V_{peak}(z/D)=1$ corresponds to the full available induced charge $-N_{e}e$.} 
\label{fig:theory}
\end{figure}

Figure~\ref{fig:theory} realizes Eq.~\eqref{eq:voutxD} for two situations. The first corresponds to a large drift velocity so that the duration of the current is short compared to the decay time of the preamplifier ($t_{e}\ll\tau_{p}$). The second corresponds to smaller drift speeds so that $t_{e}$ is equal to $\tau_{p}$. The first curve shows a linear relationship, meaning the majority of the full available induced charge given in Eq.~\eqref{eq:qinde} is preserved, while the second is strongly nonlinear for large $z/D$ due to the loss of signal from the discharge through the resistor, even as the ionization electrons continue to drift.

It is also possible to use Eq.~\eqref{eq:voutxD} to determine the drift velocity of electrons in a gas because all other terms in the equation are determined by the experimental setup. However, the sensitivity of this method is reduced when $t_{e}\ll\tau_{p}$. Large values of the drift velocity (and therefore small $t_{e}$) effectively linearize Eq.~\eqref{eq:voutxD}, thereby suppressing the $v_{e}$ dependence. This leads to a loss in sensitivity of the drift velocity on the output voltage. 

In addition to the preamplifier, a shaping amplifier is often used to increase the signal--to--noise ratio of the output signal. It is reasonable to expect that the shaping amplifier also has an effect on the shape of the voltage signals shown in Fig.~\ref{fig:theory}. For our setup, however, we have calculated that the nonlinearity in $V_{peak}(z/D)$ introduced by the amplifier is less than 5\% (see Appendix). For the remainder of this work we therefore consider only the preamplifier when determining the expected output voltage.

\section{Experimental Setup}
\label{sec:setup}
\begin{figure}[tb]
\centering
\includegraphics[width=5 in]{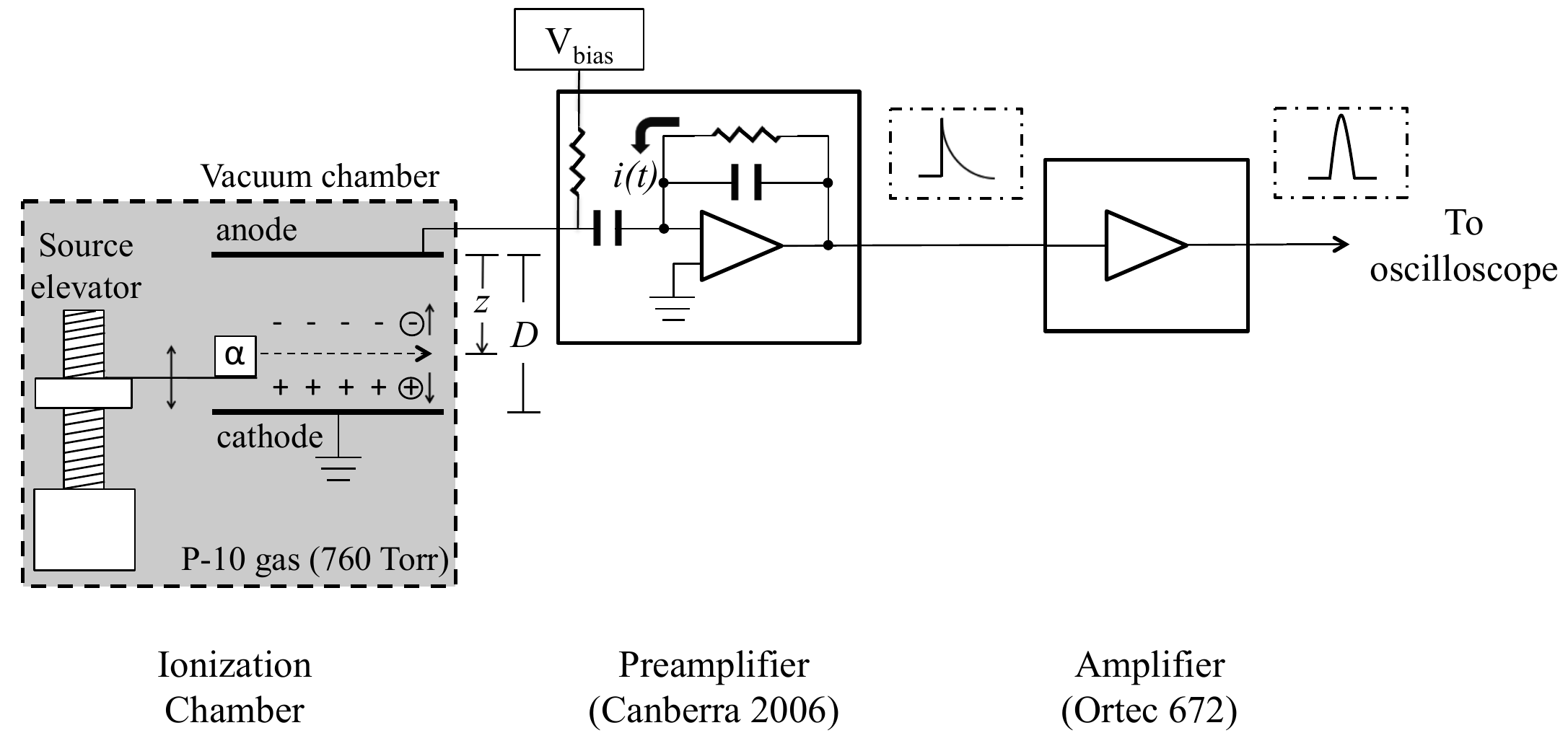}
\caption{Diagram of the full signal chain. An alpha source sits on an adjustable--height platform (see Section~\ref{sec:elevator}) and generates tracks of  ionization (dotted line) at constant $z$ inside the ionization chamber. The ionized electrons and ions drift and induce a current on the anode, which is integrated by the preamplifier to produce a voltage signal. The amplifier then shapes the output voltage pulse.}
\label{fig:sigpath}
\end{figure}

A schematic of our experimental setup is shown in Fig.~\ref{fig:sigpath}. The ionization chamber is evacuated to 10~mTorr and then back--filled with P--10 gas to 760 Torr (1~atm). The dimensions of the rectangular copper electrodes are 21~cm by 25~cm, and they are separated by 5.08~cm. In order to generate electron--ion pairs at a constant $z$, we use a collimated $^{241}$Am ionizing alpha particle source oriented parallel to the electrodes. Alpha particles emerge from the collimator at an average rate of 1~Hz. According to SRIM--2011,~\cite{srim} the range of a 5.5~MeV  $^{241}$Am alpha particle in 760 Torr of P--10 gas is 4.7 cm, so all of the alpha particles are fully stopped in the active region. The source sits approximately 3.5 cm inside of the active region in order to minimize the effects of fringe fields. In order to explore a range of values of $E/P$, we applied four voltages to the anode; the corresponding values of $E/P$ were 5.5~V~cm$^{-1}$~atm$^{-1}$, 11~V~cm$^{-1}$~atm$^{-1}$, 16.5~V~cm$^{-1}$~atm$^{-1}$, and 22~V~cm$^{-1}$~atm$^{-1}$. The source was then moved from $z$~=~0.5~cm to $z$~=~4.9~cm in intervals of 3.96~mm.
 
\subsection{Changing the Source Position}
\label{sec:elevator}
\begin{figure}[tb]
\centering
\includegraphics[width=5.5 in]{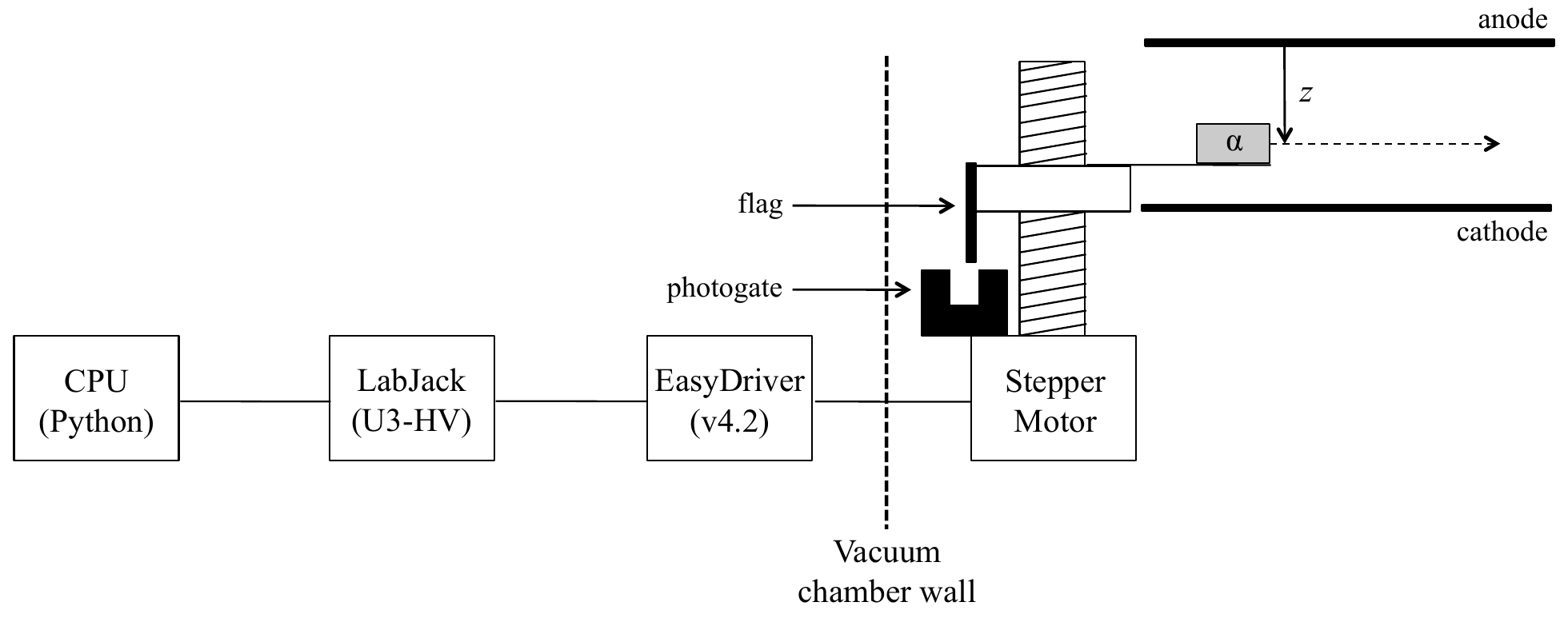}
\caption{Diagram of the elevator components (not to scale). A Python script is used to control a LabJack~U3--HV to issue motor step requests and count the number of steps. The Easy Driver provides current pulses to drive the stepper motor. An alpha source sits on the elevator inside the ionization chamber. The photogate and flag are in place to measure when the elevator has returned to its home (zero) position.}
\label{fig:labjack}
\end{figure}

\begin{figure}[tb]
\centering
\includegraphics[width=3.4 in]{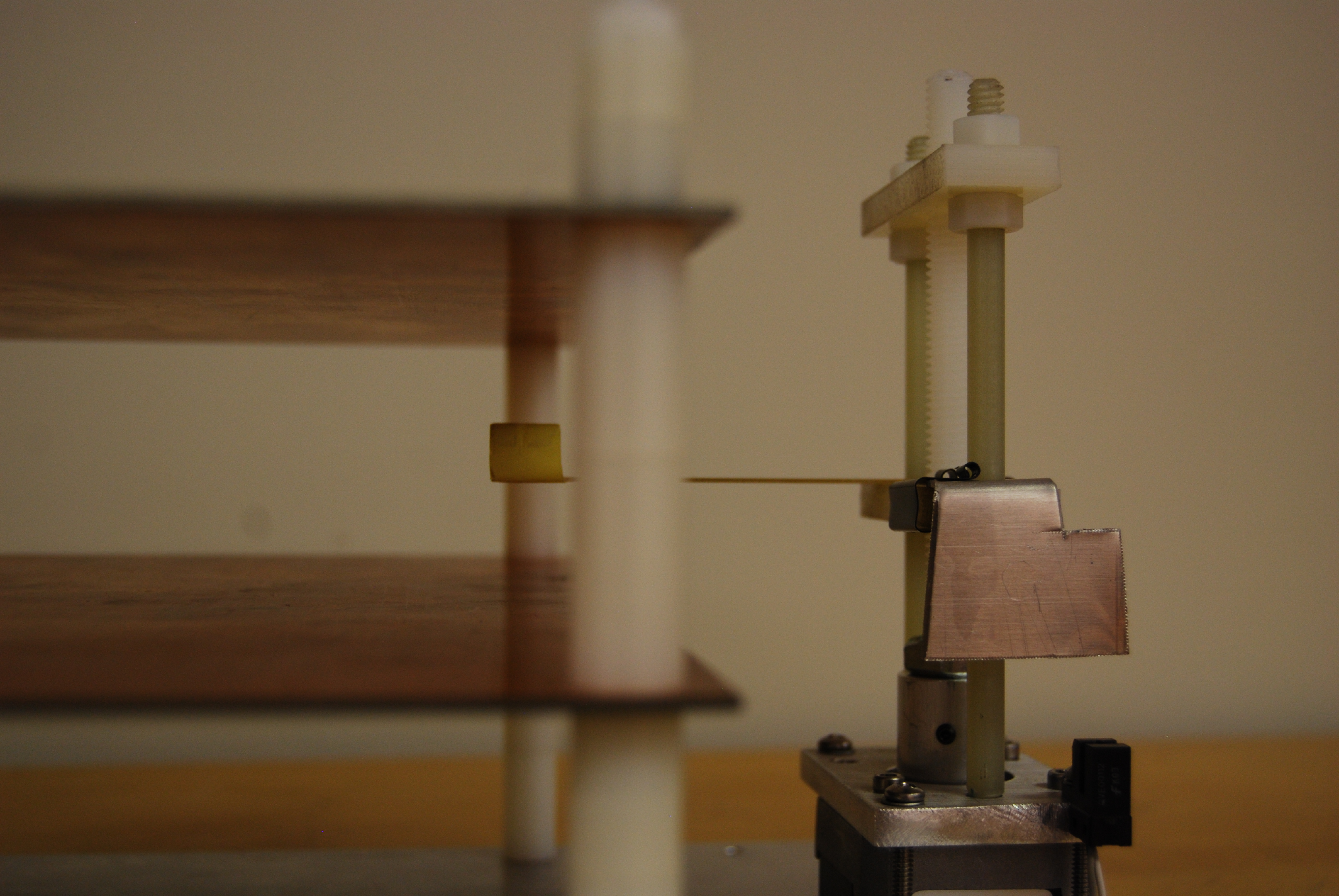}
\caption{The elevator and source in their positions in the ionization chamber. The source is housed in the yellow cylindrical collimator located between the anode (top electrode) and cathode (bottom electrode). A plastic board is used to keep the source inside the active region. The motor and elevator are seen on the right, with a metal flag attached to the elevator, and a photogate (black rectangle) attached to the motor base. The photogate and flag are used to calibrate the zero point of the drift distance variable.}
\label{fig:elevatorpics}
\end{figure}

In order to study the $z$--dependence of the measurable induced charge, we constructed a computer--controlled motorized stage (the ``elevator'') that moves the source to different distances from the anode ($z$) without the need to open the vacuum chamber. The elevator consists of a stepper motor (Mercury Motor~SM--42BYG011--25) that is driven by an EasyDriver Stepper Motor Driver (v4.2), which receives step commands from a LabJack~U3--HV multifunction data acquisition device. A threaded rod attached to the motor axle converts motor steps ($\Delta\theta$) into vertical displacements ($\Delta z$) such that $z$ changes by 0.99~mm per 1000~steps of the motor. The chamber has no viewports, so we recalibrate the home position of the motor with the use of a flag that blocks a photogate at one extreme of the motion. The flag is a rectangular piece of metal that extends beyond the edge of the elevator. In practice, the elevator position calibration is very stable in time, and the elevator returned to the correct home position without manual intervention throughout this experiment. Figure~\ref{fig:labjack} shows a schematic of the source elevator and drive system, and Fig.~\ref{fig:elevatorpics} shows images of the actual apparatus.

\subsection{Charge Readout Electronics}
\label{sec:electronics}
Using a voltage pulse generator of variable amplitude, we measure the combined gain of the preamplifier (Canberra Model 2006, with a nominal gain of 235~mV per million electron--ion pair) and the amplifier (Ortec Model 672, with a nominal gain of 100 with unipolar Gaussian signal shaping and a shaping time of 10~$\mu$s) to be 23.5~$\pm\,0.2$~V per million electron--ion pairs. The decay time of the preamplifier was measured to be 46~$\pm\,1\,\mu$s. 

\section{Results and Analysis} 
\label{sec:results} 
\begin{figure}[tb]
\centering
\includegraphics[width=3.2 in]{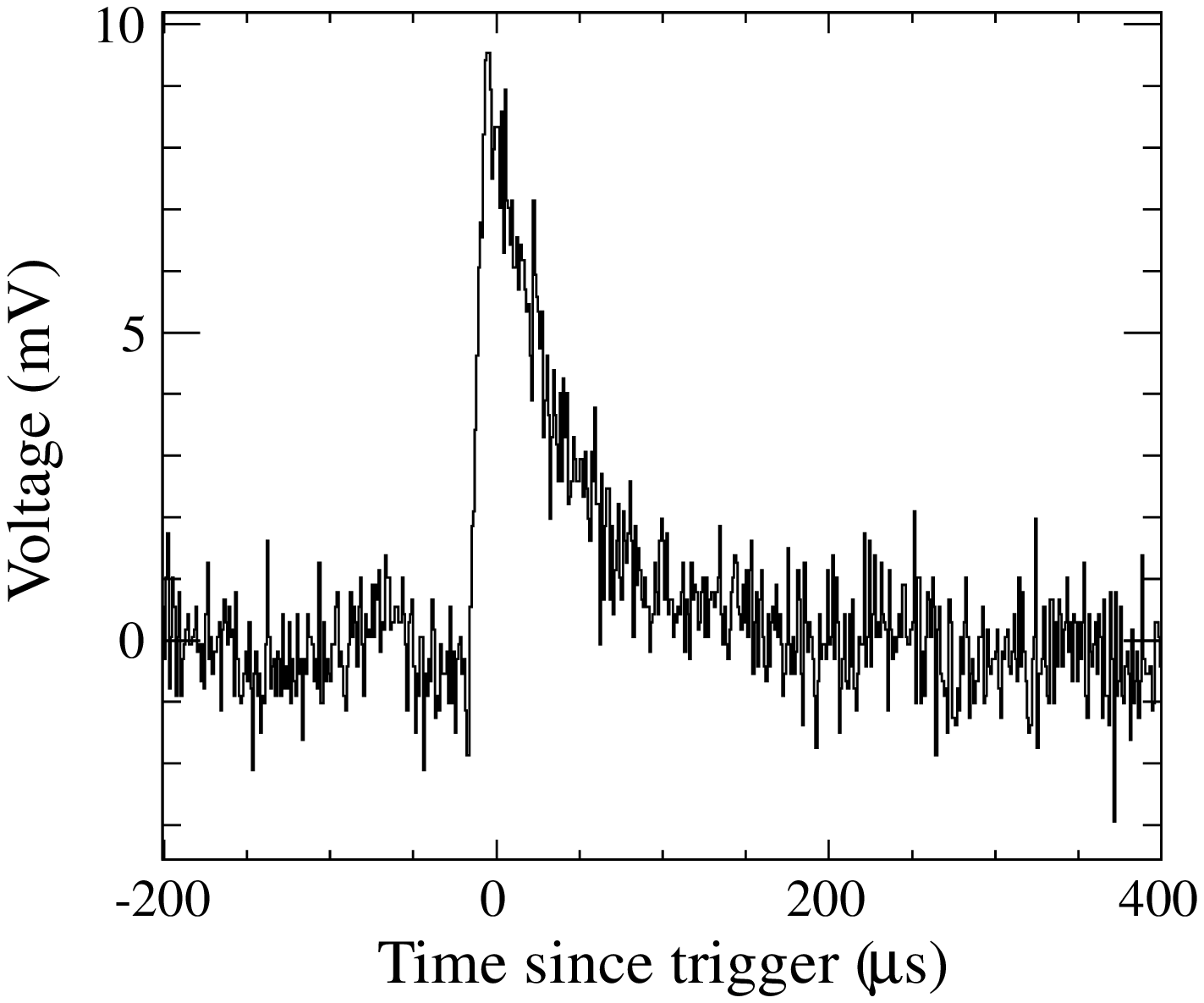} 
\includegraphics[width=3.2 in]{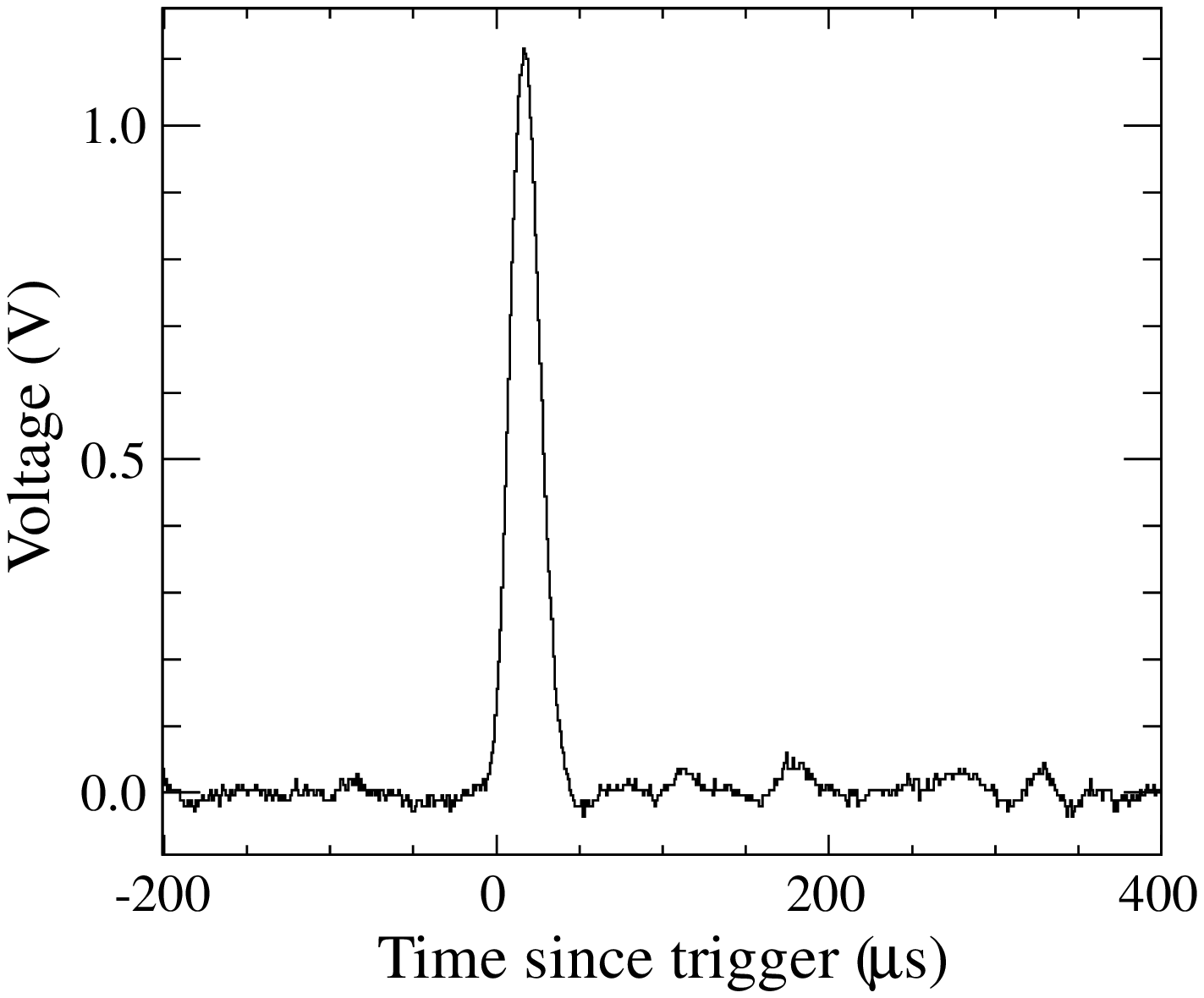}
\caption{Examples of the preamplifier and amplifier output signals at a drift distance of 2.38~cm and drift field of 5.5~V/cm/atm. The left is the output of the preamplifier, and the right is the output of the amplifier.}
\label{fig:histos}
\end{figure}

\begin{figure}[tb]
\centering
\includegraphics[width=2.8 in]{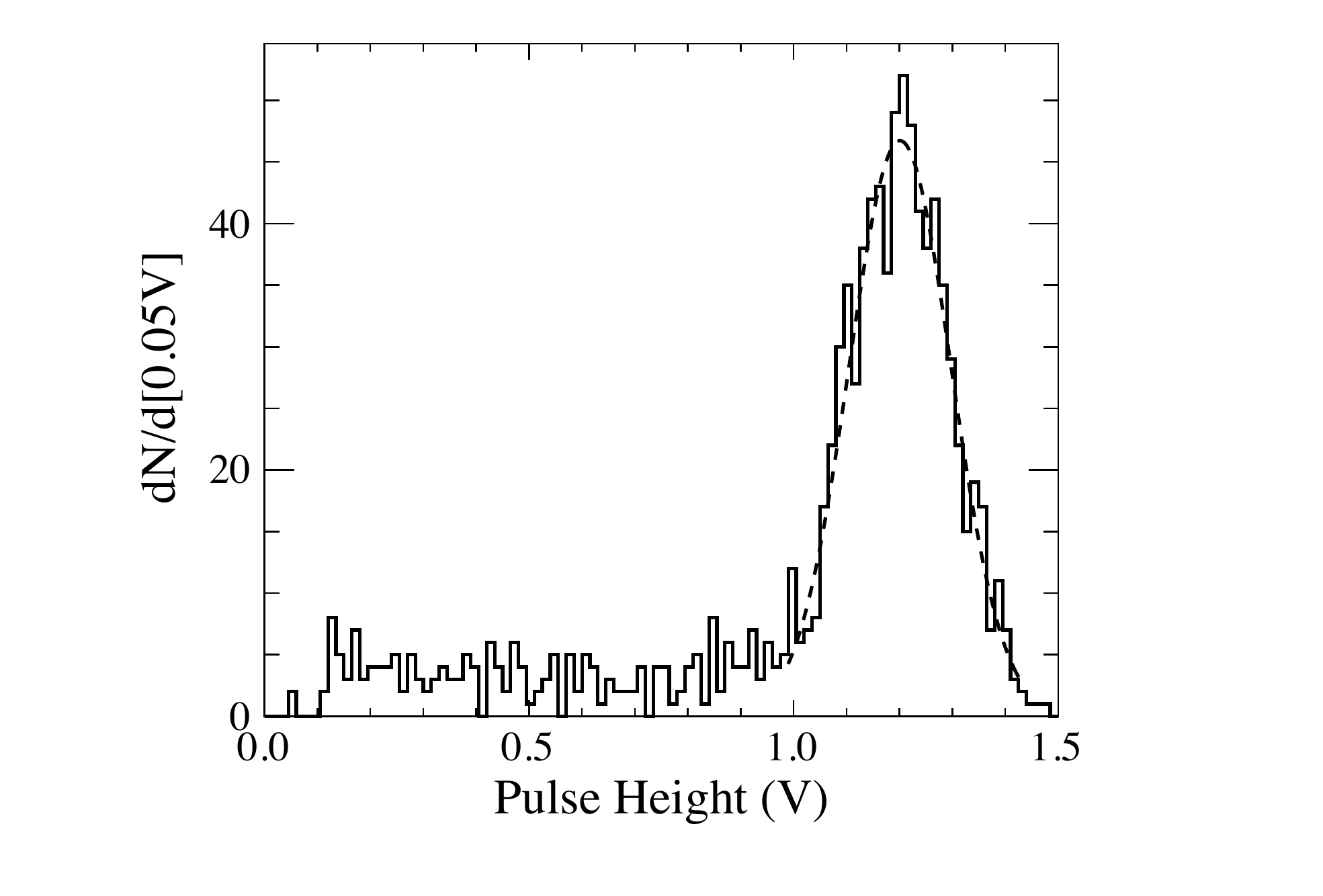}
\caption{Example pulse amplitude spectrum for a drift distance of 2.38~cm and drift field of 16.5~V/cm/atm with a Gaussian fit (dotted line).}
\label{fig:sampleph}
\end{figure}

\begin{figure}[tb]
\centering
\includegraphics[width=3.4 in]{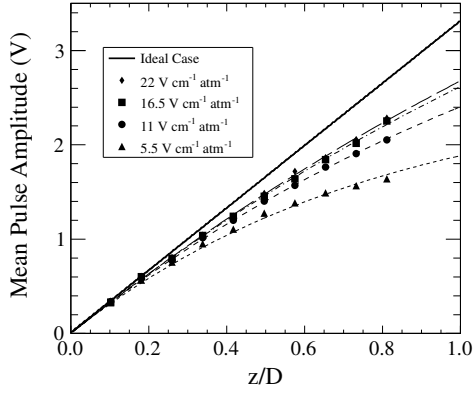}
\caption{Mean pulse amplitude vs. drift distance for four values of $E/P$. The data are shown as markers and the lines are fits of Eq.~\eqref{eq:voutxD} to each data set.}
\label{fig:driftvfigs}
\end{figure}

The output waveform from the amplifier is digitized and recorded on an oscilloscope and then transferred to a computer for analysis.~\cite{footnote2} Figure~\ref{fig:histos} shows a sample waveform from the preamplifier and the amplifier. We generate a pulse amplitude spectrum from the set of waveforms for a specific drift field ($\vec{E}$) drift distance ($z$) and determine the mean pulse amplitude from a Gaussian fit to the spectral peak (see Fig.~\ref{fig:sampleph}). The mean pulse amplitude is then plotted against the drift distance for each value of $E/P$, as shown in Fig.~\ref{fig:driftvfigs}. We fit Eq.~\eqref{eq:voutxD} to the data using our measured value of the decay time $\tau_{p}$. Because the system gain and number of electrons should be the same for all $z$, the data corresponding to the highest value of $E/P$ (22~V~cm$^{-1}$~atm$^{-1}$) is fit first in order to estimate the value of $AN_{e}$. This resulted in a value for $N_{e}$ of 0.15~million electron--ion pairs, which corresponds to an alpha particle energy of 4~MeV. 
This is consistent with expectations because 5.5~MeV alpha particles lose approximately 1~MeV in the collimator, according to SRIM--2011, and the source sits behind a thin layer of gold foil within the collimator, which accounts for the remaining 0.5~MeV energy loss. This estimated value for $AN_{e}$ is then used when fitting Eq.~\eqref{eq:voutxD} to the rest of the datasets.

Figure~\ref{fig:driftvfigs} indicates that $V_{peak}(z/D)$ increases with increasing electron drift distance, demonstrating that charge induction occurs inside the ionization chamber. If the detectable signal were due to the arrival of electrons at the anode, the output voltage would not depend on $\vec{E}$ or $z$. The accuracy of the fit of Eq.~\eqref{eq:voutxD} to the data confirms that the ion signal is negligible. The nonlinearity of $V_{peak}(z/D)$ results primarily from the $RC$ discharge in the preamplifier, and as the drift field increases (and thus $v_{e}$ increases) this nonlinearity diminishes. A faster drift velocity leads to a shorter drift time and less $RC$ discharge, and therefore a more linear relationship between drift distance and output voltage. This can be seen by comparing the data taken at 5.5~V~cm$^{-1}$~atm$^{-1}$ and 11~V~cm$^{-1}$~atm$^{-1}$. There is clear nonlinearity in the 5.5~V~cm$^{-1}$~atm$^{-1}$ data, meaning that the drift time of the electrons must be comparable to the decay time of the preamplifier. The 11~V~cm$^{-1}$~atm$^{-1}$ data show less nonlinearity, indicating that the electron drift time has decreased with an increase in the drift field (as expected). The 16.5~V~cm$^{-1}$~atm$^{-1}$ and 22~V~cm$^{-1}$~atm$^{-1}$ data are very similar, which demonstrates the fact that the output voltage becomes less sensitive to the drift velocity for large values of $v_{e}$, as discussed in Section~\ref{sec:nonideal}. The error reported for each data point in Fig.~\ref{fig:driftvfigs} is the uncertainty in the mean of the Gaussian fit to the pulse amplitude spectrum. In all cases, the error bars are smaller than the size of the data markers.

It should be noted that the loss of electrons through attachment to other neutral gas molecules in the chamber can also cause a decrease in the total output voltage and mimic the nonlinearity that comes from the drift velocity. In order to quantify the contribution of this, we held the source at a constant value of $z/D$ but varied the value of $D$ by a factor of two and measured the mean pulse amplitude. From this, we determined that attachment is responsible for no more than 16\% of the decrease in the output voltage signal from the highest value of $E/P$ to the lowest value. Therefore, the majority of the observed nonlinearity comes from the preamplifier, as expected. However, it is possible that attachment still impacts the signal. In Fig.~\ref{fig:driftvfigs}, a linear ideal case is displayed with the data in order to show the overall departure from linearity. Even the highest values of $E/P$ have not produced a completely linear output signal, which indicates that attachment may be decreasing the overall induced charge for all data sets. This discrepancy is consistent with the 16\% loss of signal mentioned above. 

We are not aware of measurements of the electron drift speed $v_{e}$ in P--10 gas for the low values of $E/P$ used here ($E/P <$ 22~V~cm$^{-1}$~atm$^{-1}$). However, this experiment is sensitive to the drift velocity through the nonlinearity of $V_{peak}(z/D)$ vs. $z/D$ and therefore provides a way to measure $v_{e}$ for low $E/P$. We can see that the drift velocity grows with $E/P$ as expected from Eq.~\eqref{eq:ve}, and  from the fact that a faster drift velocity corresponds to a shorter drift time, meaning that less signal discharges through the preamplifier feedback resistor and therefore the data is increasingly linear. 

We choose not to report any drift velocity measurements with certainty here because electron drift speeds can be dependent on gas composition and impurities,~\cite{asharma} a parameter only loosely controlled in this experiment. ÊAlso, the non--linearity in the $V_{out}(z/D)$ data (Fig.~\ref{fig:driftvfigs}) results not only from the electron drift velocity but also from electron attachment onto electronegative gas impurities. ÊTherefore, our measurement of the electron drift speed would likely underestimate the electron drift speed. ÊA similar setup, but with better control of gas purity, could be used to measure the electron drift speed. ÊAdditionally, we note that there are several more direct ways to measure the electron drift speed in gases.~\cite{colas}$^{,}$~\cite{pushkin}$^{,}$~\cite{christophorou}	

\section{Conclusions}
\label{sec:conclusions}
We have shown that an ionization chamber and a collimated alpha particle source on an adjustable stage are effective in demonstrating the phenomenon of charge induction. By measuring the mean pulse amplitude vs. drift distance it is possible to study gas ionization, the relative motion of electrons and ions in a gas, and the dependence of that motion on other variables such as drift field and distance. This experiment is also useful for studying the effects of electronics on measured signals. This work can be extended to measure the $W$--value of various detector gases~\cite{schaefer}$^{,}$~\cite{wolfe} (essentially by fitting for $N_{e}$), or to determine the $z$--coordinate of an ionization event inside the chamber.~\cite{whitepaper}$^{,}$~\cite{drift}

\appendix*

\section{Amplifier Effect Calculation}
\label{sec:ampappendix}

Although the calculation of the convolution of a rectangular current pulse and an exponentially decaying preamplifier response is readily available (e.g., Refs.~\onlinecite{blum} and \onlinecite{sharma}), we are not aware of such a calculation with both a preamplifier and the amplifier. We include both calculations here.~\cite{appendixfootnote}

\subsection{Output of Preamplifier Only}
A typical preamplifier has an impulse response of 
\begin{equation}
\label{eq:preamptrans}
h_{p}(t)=Ae^{-t/\tau_{p}} \quad \text{for} \,\,\, t \ge 0,
\end{equation}
where $A=\dfrac{1}{C_{f}}$ is the gain of the preamplifier and $\tau_{p}=R_{f}C_{f}$ is the decay time of the preamplifier. The input signal in this case is the induced current on the anode [see Eq.~\eqref{eq:ie1}], which is constant for a time equal to the drift time of the electrons $t_{e}$. The convolution of $i_{e}(t)$ with $h_{p}(t)$ then represents the preamplifier output voltage, $V_{out,p}(t)$, or
\begin{align}
\label{eq:voutpreamp}
V_{out,p}(t) = i_{e}(t) \otimes h_{p}(t) = 
\begin{dcases}
A\tau_{p}\frac{N_{e}e\,v_{e}}{D}\left(1-e^{-t/\tau_{p}} \right) & 0 < t <  t_{e} \\
A\tau_{p}\frac{N_{e}e\,v_{e}}{D}\left(e^{(-t+t_{e})/\tau_{p}}-e^{-t/\tau_{p}} \right) & t > t_{e}.
\end{dcases}
\end{align}

\subsection{Output of Preamplifier and Amplifier}
\begin{figure}[tb]
\centering
\includegraphics[width=3.4 in]{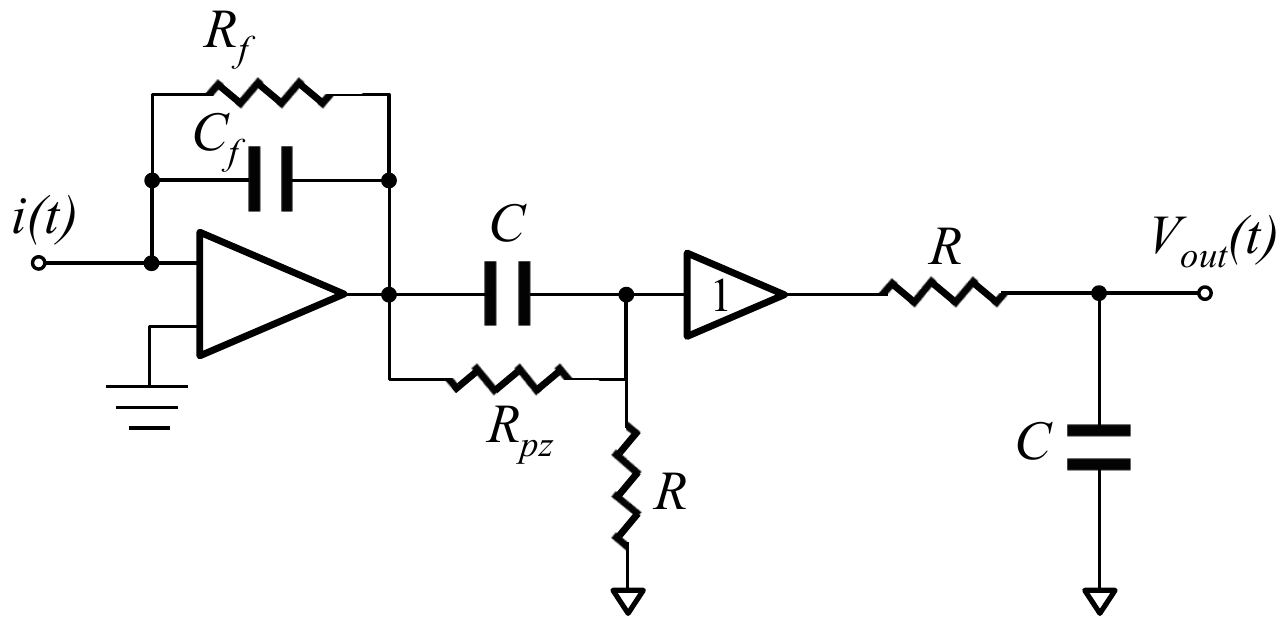}
\caption{Circuit diagram of the preamplifier and amplifier with pole-zero cancellation.} 
\label{fig:pzpreampamp}
\end{figure}

When both a preamplifier and an amplifier are used, the output signal is the convolution of the impulse response of the preamplifier and the amplifier, again convolved with the induced current $i_{e}(t)$. A diagram of the entire circuit is shown in Fig.~\ref{fig:pzpreampamp}. We first determine the total impulse response of the combined preamplifier and amplifier. Equation~\eqref{eq:preamptrans} gives the impulse response for the preamplifier, and the impulse response for a CR--RC amplifier with pole zero cancellation is~\cite{sharma}
\begin{equation}
\label{eq:amptrans}
h_{a}(t)=B\frac{e^{\frac{-t}{\tau_{a}}}}{\tau_{a}^2}\left[\tau_{a}+\tau_{z}\left(e^{\frac{-t}{\tau_{z}}}-1 \right) \right],
\end{equation}
where $\tau_{a}=RC$ is the shaping time of the amplifier, $B$ is the voltage gain of the amplifier, and $\tau_{z}=R_{pz}C$ is the timescale for the pole-zero cancellation circuit. The pole--zero adjustment functions to eliminate undershoot in the output voltage waveform. This is achieved when $\tau_{z}=\tau_{p}$, and here we assume this adjustment has been made. The total transfer function of the preamplifier and amplifier is then
\begin{equation}
\label{eq:htot}
h_{total}(t)=h_{p}(t) \otimes h_{a}(t)=AB\left(\frac{\tau_{p}}{\tau_{a}}\right)e^{\frac{-t}{\tau_{a}}}\left (1-e^{\frac{-t}{\tau_{p}}} \right).
\end{equation} 

Given the system's impulse response, we can determine the output signal $V_{out,p+a}(t)$ of the amplifier by convolving Eq.~\eqref{eq:htot} with the input current signal in Eq.~\eqref{eq:ie1}, giving 
\begin{align}
\label{eq:voutamp}
V_{out,p+a}(t)=
\begin{dcases}
	AB\tau_{p}\frac{N_{e}e\,v_{e}}{D}\left[\frac{\tau_{p}}{\tau_{p}+\tau_{a}}-e^{\frac{-t}{\tau_{a}}}\left(1-\frac{\tau_{p}}{\tau_{p}+\tau_{a}}e^{\frac{-t}{\tau_{p}}}\right)\right] & 0 < t <  t_{e} \\
	AB\tau_{p}\frac{N_{e}e\,v_{e}}{D}\bigg[e^{-t/\tau_{a}}\left(e^{t_e/\tau_{a}}-1\right) \\ \hspace{1 in} - \frac{\tau_{p}}{\tau_{p}+\tau_{a}}e^{-t\left(\frac{1}{\tau_{a}}+\frac{1}{\tau_{p}}\right)}\left(e^{t_e\left(\frac{1}{\tau_{a}}+\frac{1}{\tau_{p}}\right)}-1\right)\bigg] & t > t_{e}	.
\end{dcases}
\end{align}

\subsection{Peak Value Ratio Calculation}
\begin{figure}[tb]
\centering
\includegraphics[width=3.4 in]{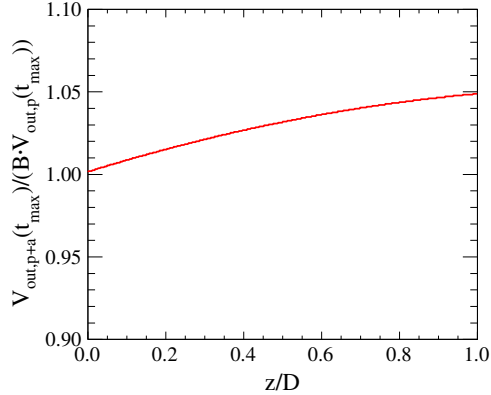}
\caption{Ratio of the maximum values of $V_{out,p+a}$ and $V_{out,p}$ as a function of $z/D$.} 
\label{fig:voutratiograph}
\end{figure}

In order to determine the magnitude of the effect that the amplifier has on the voltage output signal, we calculated the ratio of Eq.~\eqref{eq:voutamp} to Eq.~\eqref{eq:voutpreamp} at $t_{max}$, where $t_{max}$ is the time of maximum amplitude. Note that for $V_{out,p}(t)$, $t_{max}=t_{e}$, but for $V_{out,p+a}$, $t_{max} > t_{e}$. We used the nominal values of $\tau_{p}$ and $\tau_{a}$ for our experimental setup ($\tau_{p}$~=~50~$\mu$s, $\tau_{a}$~=~10~$\mu$s), and wrote $t_{e}$ in terms of $z/D$. As shown in Fig.~\ref{fig:voutratiograph}, the resulting curve deviates from unity by a maximum of 5\%, meaning that the amplifier does not contribute significantly to the nonlinearity of $V_{out}(z/D)$. Instead, the nonlinearity in $V_{out}(z/D)$ is dominated by the interplay between the electron drift time and the preamplifier decay time, which is why we have chosen to ignore the amplifier in Section~\ref{sec:nonideal}.

\begin{acknowledgments}
The authors would like to acknowledge the Dark Matter Time Projection Chamber group for their support and Igal Jaegle for assistance with the MAGBOLTZ simulation. The authors would also like to thank Rich Willard and Rob Cunningham at Bryn Mawr College for their assistance with the design and fabrication of the experimental apparatus. J.B.R.B. acknowledges the support of the Pappalardo Fellowships in Physics at MIT. Lastly, the authors would like to thank the anonymous reviewers for strengthening this work with their valuable comments.
\end{acknowledgments}


\begin{thebibliography}{5}
\bibitem{footnote0}The motion of the electron also produces a current of the opposite sign on the bottom plate, the magnitude of which corresponds to the distance between the electron and the bottom plate.
\bibitem{footnote1}There are many other ways for a particle to lose energy while moving through a medium; see Refs.~\onlinecite{ahlen} and~\onlinecite{pdg}.
\bibitem{footnote2}Note that a multichannel analyzer could also be used to create a pulse height spectrum.
\bibitem{appendixfootnote}We have chosen to present these calculations in terms of convolutions and impulse responses rather than Laplace Transforms and transfer functions to avoid the unnecessary discussion of Laplace Transforms and their inverses.
\end{thebibliography}
\end{document}